\documentclass[preprint,review,12pt]{elsarticle}

\usepackage[utf8]{inputenc}
\usepackage{graphicx}
\usepackage{amssymb}
\usepackage{amsmath}
\usepackage{amsfonts}
\usepackage{color}
\usepackage{hyperref}
\usepackage[english]{babel}
\usepackage{subfigure}
\usepackage{algorithm}
\usepackage{algorithmic}
\usepackage{multicol}
\usepackage{multirow}
\usepackage{tabularx,colortbl}
\usepackage{lineno}

\usepackage{tikz}
\usetikzlibrary{calc, shapes, backgrounds}
\tikzset{
  head/.style = {fill = orange!80,
                 label = center:\textsf{\footnotesize{MV}}},
  tail/.style = {fill = orange!80,
                 label = center:\textsf{}}
}

\biboptions{sort&compress}
\newtheorem{thm}{Def.}

\definecolor{lgray}{rgb}{0.9,0.9,0.9}

\journal{Int. Journal of Electrical Power \& Energy Systems}

\definecolor{Orange}{rgb}{1,0.64,0}
\definecolor{lgray}{rgb}{0.9,0.9,0.9}

\newcommand{\argmin}{\operatornamewithlimits{arg\ min}}

\makeindex

\begin{document}

\begin{frontmatter}
\title{On the impact of topological properties of smart grids in power losses optimization problems}
\author[address1]{Francesca Possemato}
\ead{francesca.possemato@uniroma1.it}
\author[address1]{Maurizio Paschero}
\ead{maurizio.paschero@uniroma1.it}
\author[address2]{Lorenzo Livi\corref{cor1}}
\ead{llivi@scs.ryerson.ca}
\ead[url]{https://sites.google.com/site/lorenzlivi/}
\author[address1]{Antonello Rizzi}
\ead{antonello.rizzi@uniroma1.it}
\ead[url]{http://infocom.uniroma1.it/~rizzi/}
\author[address2]{Alireza Sadeghian}
\ead{asadeghi@ryerson.ca}
\ead[url]{http://www.scs.ryerson.ca/~asadeghi/}
\cortext[cor1]{Corresponding Author}

\address[address1]{Dept. of Information Engineering, Electronics, and Telecommunications, SAPIENZA University of Rome, Via Eudossiana 18, 00184 Rome, Italy}
\address[address2]{Dept. of Computer Science, Ryerson University, 350 Victoria Street, Toronto, ON M5B 2K3, Canada}

\begin{abstract}
Power losses reduction is one of the main targets for any electrical energy distribution company.
In this paper, we face the problem of joint optimization of both topology and network parameters in a real smart grid. We consider a portion of the Italian electric distribution network managed by the ACEA Distribuzione S.p.A. located in Rome.
We perform both the power factor correction (PFC) for tuning the generators and the distributed feeder reconfiguration (DFR) to set the state of the breakers.
This joint optimization problem is faced considering a suitable objective function and by adopting genetic algorithms as global optimization strategy. 
We analyze admissible network configurations, showing that some of these violate constraints on current and voltage at branches and nodes.
Such violations depend only on pure topological properties of the configurations.
We perform tests by feeding the simulation environment with real data concerning hourly samples of dissipated and generated active and reactive power values of the ACEA smart grid.
Results show that removing the configurations violating the electrical constraints from the solution space leads to interesting improvements in terms of power loss reduction.
To conclude, we provide also an electrical interpretation of the phenomenon using graph-based pattern analysis techniques.
\end{abstract}
\end{frontmatter}


\section{Introduction}
\label{sec:intro}
In recent years, the global warming has led people and companies to demand for cleaner energy suppliers. Thus, more and more electricity is generated from alternative and heterogeneous sources: wind, solar, biofuel, and geothermal plants. This phenomenon is called Distributed Generation (DG) \cite{ghosh2010optimal,moradi2012combination,kumar2014reliability}.
A Smart Grid (SG) \cite{5535240} constitutes the improvement of a traditional electrical distribution system, which is conceived to overcome the problem of the wide diffusion and high penetration of DGs.
A SG can be seen as an intelligent network able to integrate all users (i.e., producers and consumers) with the ultimate purpose of distributing the electrical power in a safe, efficient, and sustainable fashion \cite{Dahu_2011,Farhangi_2010,venayagamoorthy2011dynamic,gentile2014reactive,pagani2014power}.
With the advent of SGs, the customers of electrical networks become also energy suppliers and the load flow in distribution feeders becomes bidirectional.
Moreover, a large number of sensors are installed on the network to obtain a complete information on the instantaneous status of the infrastructure -- information that could be exploited for predicting faults \cite{occ_sg_enricods__arxiv,zhang2011fault,saha2011fault}.

The reduction of power losses is one of the main objectives of energy electrical distribution companies.
In the literature \cite{Chandramohan_2010} it is possible to identify two mainstream approaches: \textit{Power Factor Correction} (PFC) and \textit{Distributed Feeder Reconfiguration} (DFR).
The PFC tries to reduce the amount of reactive power present in the network in order to (i) minimize the Joule losses, (ii) increase the capacity of the network, and (iii) increase the quality of service.
The DFR, instead, relies on switching a certain number of breakers, physically modifying the topological structure of the network and improving its operating conditions.
In doing so, operational constraints on the network must be satisfied, such as ensuring that no loops are formed and the totality of the loads are supplied.
Altering the network configuration affects the power losses and relieves overload in the network; thus the DFR problem can be conceptualized as the task of choosing the status of the network breakers resulting in the configuration with minimum power losses, yet still satisfying the operational constrains.
The main drawback of DFR is that it results in a complex combinatorial optimization problem, since the status of the switches is non-differentiable. This makes the optimization problem related to DFR very hard to solve.
Many researchers have proposed interesting solutions in the past. \citet{Civanlar_1988} propose a heuristic method based on a formula that expresses the change in losses between network configurations before and after the reconfiguration process. Moreover in the paper the authors suggest a method for filtering configurations that yield lower reduction in power losses.
Other heuristic approaches are presented in Refs. \cite{Nara_1992,Merlin_1975}. All such mentioned methods consist in approximate solutions or local optimum of the respective optimization problems. To overcome this issue, \citet{McDermott_1999} use a genetic algorithm providing a good compromise between computational burden and quality of the optimization result.
More recently, novel meta-heuristic methods based on evolutionary optimization algorithms are introduced for the same context, showing good experimental results \cite{olamaei2008application,niknam2009efficient,malekpour2013multi,mazza2014optimal,rao2013power}.

One of the main technical difficulties in dealing with the DFR problem using evolutionary algorithms is the so-called ``radiality constraint''.
The fulfillment of this constraint makes inappropriate most of the network configurations achievable by switching the available breakers.
For this reason, it is necessary to conceive a procedure able to select, among all possible configurations, those fulfilling the radiality constraint.
A first solution to this problem for the SG of ACEA has been proposed by \citet{Storti_2014}.
The authors conclude that due to the high complexity of the DFR problem, a desirable method should be able to reduce as much as possible the solution space, eliminating undesirable switching options a priori. Such configurations are critical with respect to both topological and electrical constraints. 
In \cite{caschera_2014}, the authors propose a heuristic method to compare the admissible network configurations in a purely topological manner, facilitating the optimization algorithm in finding the desired solution.
All such studies highlight the need to identify undesirable network configurations, in order to reduce the required convergence time for the faced optimization problem. In fact, undesirable configurations, which cause violations of one or more electrical constraints, introduce a significant and unnecessary increase of the computational demand for the simulation.
This is due to the fact that the solver tries to perform PFC and DFR using also configurations that are intrinsically critical for particular power loads profiles.

Concerning other works dealing with the Joule losses minimization problem on real Smart Grids, in Ref. \cite{Mahdad_2008} a genetic algorithm with fuzzy logic rules is employed to face the optimal power flow problem in the Algerian electric network as a Flexible AC Transmission System, while in Ref. \citep{Amrane_2014} is proposed a particle swarm  optimization method for solving the optimal reactive power dispatch (ORPD) problem on the Algerian electric power system. Although not concerning the Joule losses minimization problem, it is worth to cite the work of \citet{Corsi_2004}, where the hierarchical voltage control system presently applied on the Italian transmission grid is described in details.
Moreover, in Ref. \cite{Senac_2014} the use of capacitors and static reactive power compensators to ensure the voltage stability on the electrical network serving the South-West region of France is studied, aiming to develop an advanced system for the control of the reactive power compensation.

In our previous works \cite{Storti_2013,Possemato_2013,Storti_2013_b}, we faced both the problems of PFC and DFR over a portion of the ACEA electrical grid (ACEA is the company managing the entire distribution grid of Rome, Italy), using genetic algorithms as optimization strategy.
In this paper we elaborate over such studies by first analyzing how undesirable configurations affect the optimization process in terms of both quality of the optimization result and running time of the optimization procedure.
Then we give an interpretation of the results from an electrical point of view, exploiting a graph-based pattern analysis technique. Notably, we study two prototype graphs representing two classes of typical network configurations identified in our data.
Our work is framed in the research area concerning the application of genetic algorithms to face the joint PFC and DFR problem. As concerns similar works, in Refs. \cite{Kalantar_2006}, \cite{Madeiro_2011}, \cite{Diaz_2003} active power losses minimization is faced by simultaneous capacitor placement and feeder reconfiguration by genetic algorithms. With respect to our approach, PFC is solved by installing capacitor banks to compensate the losses produced by reactive currents, while in the SG of ACEA the PFC problem is faced by regulating the power factors of the distributed generators in the network.
In fact, the telecontrol system of ACEA is capable to modify remotely the set points of each renewable energy generator. In Ref. \cite{Singh_2011} the reader can find an interesting review on optimal placement approaches of DGs in power systems for optimizing different objective functions (such as power losses minimization), including some algorithms based on evolutionary computation and swarm intelligence (e.g., genetic  algorithms,  ant colony optimization, particle  swarm optimization).
Finally, we stress that in the SG of ACEA the position of the DGs cannot be relocated, since we are dealing with a real network whose physical characteristics cannot be changed.

The paper is structured as follows. In Section \ref{sec::ProblForm}, we first describe the optimization problem (Section \ref{sec::opt_procedure}) and then we introduce the problem of admissibility for a network configuration (Section \ref{sec::AdmNetConf}). Section \ref{sec:acea_opt} provides the essential technical details (in Section \ref{sec:Acea_SG}) of the electrical distribution network under analysis (the ACEA SG) and the related power loss optimization problem (in Section \ref{sec::Opt_Prob_ACEA}). In Section \ref{sec:dis_FF} we analyze the admissible configurations and we introduce the concept of constraint compliant configuration.
In Section \ref{sec::ExpRes} we present and discuss the results of the optimization, providing also an electrical interpretation aimed at providing a justification for them.
Finally, in Section \ref{sec:conclusions} we draw the conclusions pointing at the future directions.
\section{Power Loss Minimization Problem}
\label{sec::ProblForm}
In this paper, we consider the joint PFC and DFR problem for minimum power losses, satisfying constraints on nodes voltage and branches current as well as system operating constraints.

\subsection{Optimization Problem}
\label{sec::opt_procedure}

In this section we formulate the problem of active power losses minimization in SGs through PFC and DFR.
The problem consists in finding the optimal network parameters and the topological configuration that minimize the value of the power losses in the network, considering the constraints imposed on voltages and currents due to safety or quality of service issues as well as physical topological constraints.
Consider an admissible set $E$ of the network parameters and a suitable cost function $J:E\rightarrow\mathbb{R}$ that associates a real number to each element in $E$. Formally, the problem consists in minimizing the function $J$ in $E$. 
Mathematically, we can express the cost function $J \in [0,1)$ as follows:
\begin{equation}
J(\mathbf{k}) = \frac{P_{\mathrm{loss}}(\mathbf{k})}{P_{\mathrm{gen}}(\mathbf{k})}=\frac{P_{\mathrm{gen}}(\mathbf{k})-P_{\mathrm{load}}}{P_{\mathrm{gen}}(\mathbf{k})},
\label{eq::J}
\end{equation}
where $\mathbf{k}\in E$ represents an instance of the network parameters, $P_{\mathrm{gen}}(\mathbf{k}) \in [P_{\mathrm{load}}, \infty)$ is the total power generated by all sources, $P_{\mathrm{load}}$ is the total power absorbed by the loads, and their difference $P_{\mathrm{loss}}(\mathbf{k})$ represents the total power losses in the network.
Notice that in the present formulation of the problem, $P_{\mathrm{load}}$ is independent by the network parameters $\mathbf{k}$ because ACEA can provide only a description of the loads based on the profiles of active and reactive power measured by the meters installed in the network.

Let us consider a generic SG characterized by $n$ real parameters, $m$ integer parameters, and $p$ nominal parameters. We can express the domain of the ordinal parameters as:
\begin{equation}
\label{eq::A'}
A^{\mathrm{ord}} = \left\{ \mathbf{k}^{\mathrm{ord}} \in \mathbb{R}^{n} \times \mathbb{Z}^{m} \ \ :\  \mathbf{k}^{\mathrm{ord}}_{\mathrm{min}} \leq \mathbf{k}^{\mathrm{ord}} \leq \mathbf{k}^{\mathrm{ord}}_{\mathrm{max}} \right\},
\end{equation}
in which $\mathbf{k}^{\mathrm{ord}}_{\mathrm{min}}$ and $\mathbf{k}^{\mathrm{ord}}_{\mathrm{max}}$ represent the vectors of the minimum and maximum values of the network ordinal parameters, $\mathbf{k}^{\mathrm{ord}}$. Concerning the nominal parameters, $\mathbf{k}^{\mathrm{nom}}$, the domain is a set $A^{\mathrm{nom}}$ of all possible admissible elements for such parameters:
\begin{equation}
\label{eq::A''}
A^{\mathrm{nom}} = \left\{ \mathbf{k}^{\mathrm{nom}} \in \mathbb{X}_{1} \times \cdots  \times \mathbb{X}_{p} \right\},
\end{equation}
in which $\mathbb{X}_i$ is a generic nominal set with $i\in\{1,...,p\}$.
The overall domain $A$ is defined as $A = A^{\mathrm{ord}} \times A^{\mathrm{nom}}$; accordingly $\mathbf{k} = [\mathbf{k}^{\mathrm{ord}},  \mathbf{k}^{\mathrm{nom}}]\in A$.
In order to be valid, a solution $\mathbf{k}$ must satisfy the constraints on voltages and currents defined below:
\begin{equation}
\label{eq::BC}
\begin{aligned}
B &= \big\{ \mathbf{k} \in A : V_{i}^{\mathrm{min}} \leq V_i(\mathbf{k}) \leq V^{\mathrm{max}}_i ,i=1,...,N \big\} \\
C &=  \big\{ \mathbf{k} \in A : |I_j(\mathbf{k})|\leq I^{\mathrm{max}}_j, j=1,...,R \big\},
\end{aligned}
\end{equation}
where $V_i(\mathbf{k})$ is the voltage magnitude of the $i$-th node for a fixed instance of parameters $\mathbf{k}$, $N$ is the total number of nodes, $V^{\mathrm{min}}_i$ and $V^{\mathrm{max}}_i$ are the voltage limits for the $i$-th node, while $|I_j(\mathbf{k})|$ represents the current magnitude of the $j$-th branch for a particular instance of parameters $\mathbf{k}$, $R$ the number of branches, and finally $I^{\mathrm{max}}_j$ the current upper bound for the $j$-th branch.
The definitions given above allow to define the admissible set $E$ as follows:
\begin{equation}
E = A \cap B \cap C .
\label{eq::E}
\end{equation}

Since it is not practically possible to derive expression \eqref{eq::J} in closed-form as a function of $\mathbf{k}$, in the following we will employ a ``standard'' GA (a well-known derivative-free approach) as global optimization algorithm.
Moreover, since it is also not practically possible to derive closed-forms for $V_i(\mathbf{k})$ and $I_j(\mathbf{k})$ in \eqref{eq::BC}, we introduce a new function $\Gamma(\mathbf{k})$ used in the optimization procedure as a measure of the violation of voltage and current constraints.
Therefore, the constrained optimization problem defined above is actually faced by defining the objective function as a convex combination of the following two conflicting terms:
\begin{equation}
F(\mathbf{k}) =  \alpha J(\mathbf{k}) + (1-\alpha) \Gamma(\mathbf{k}),
\label{eq::alpha}
\end{equation}
where $\alpha\in[0, 1]$ is a parameter used to adjust the relative weight of the power losses term, $J(\mathbf{k})$, over the constraints violation term, $\Gamma(\mathbf{k})$.
Thus, our objective becomes minimizing the function $F(\mathbf{k})$ in the domain $A$ defined through \eqref{eq::A'} and \eqref{eq::A''}, instead of $J(\mathbf{k})$ \eqref{eq::J} in $E$ \eqref{eq::E}.
Note that \eqref{eq::alpha} is meaningful if and only if both $J(\mathbf{k})$ and $\Gamma(\mathbf{k})$ vary in the same range, otherwise the optimization problem is not well-posed and it is not guaranteed that minimizing $F(\mathbf{k})$ in the domain $A$ gives approximately the same result as minimizing $J(\mathbf{k})$ \eqref{eq::J} in $E$ \eqref{eq::E}.

\begin{figure}[!thbph]
\centering
\begin{tabular}{c}
\includegraphics[viewport=0 0 1289 599,scale=.25]{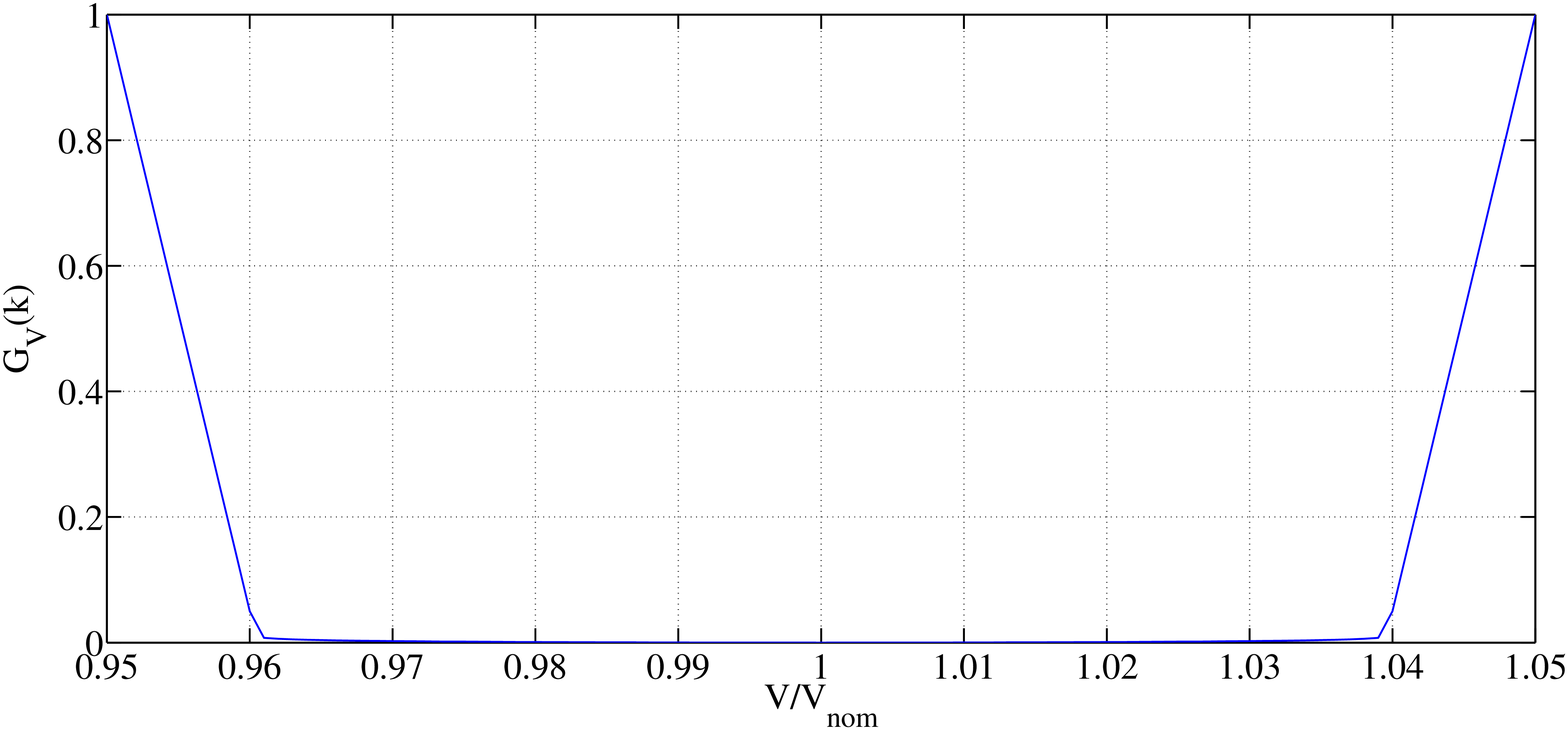} \\
(a) \\
\includegraphics[viewport=0 0 1262 599,scale=.25]{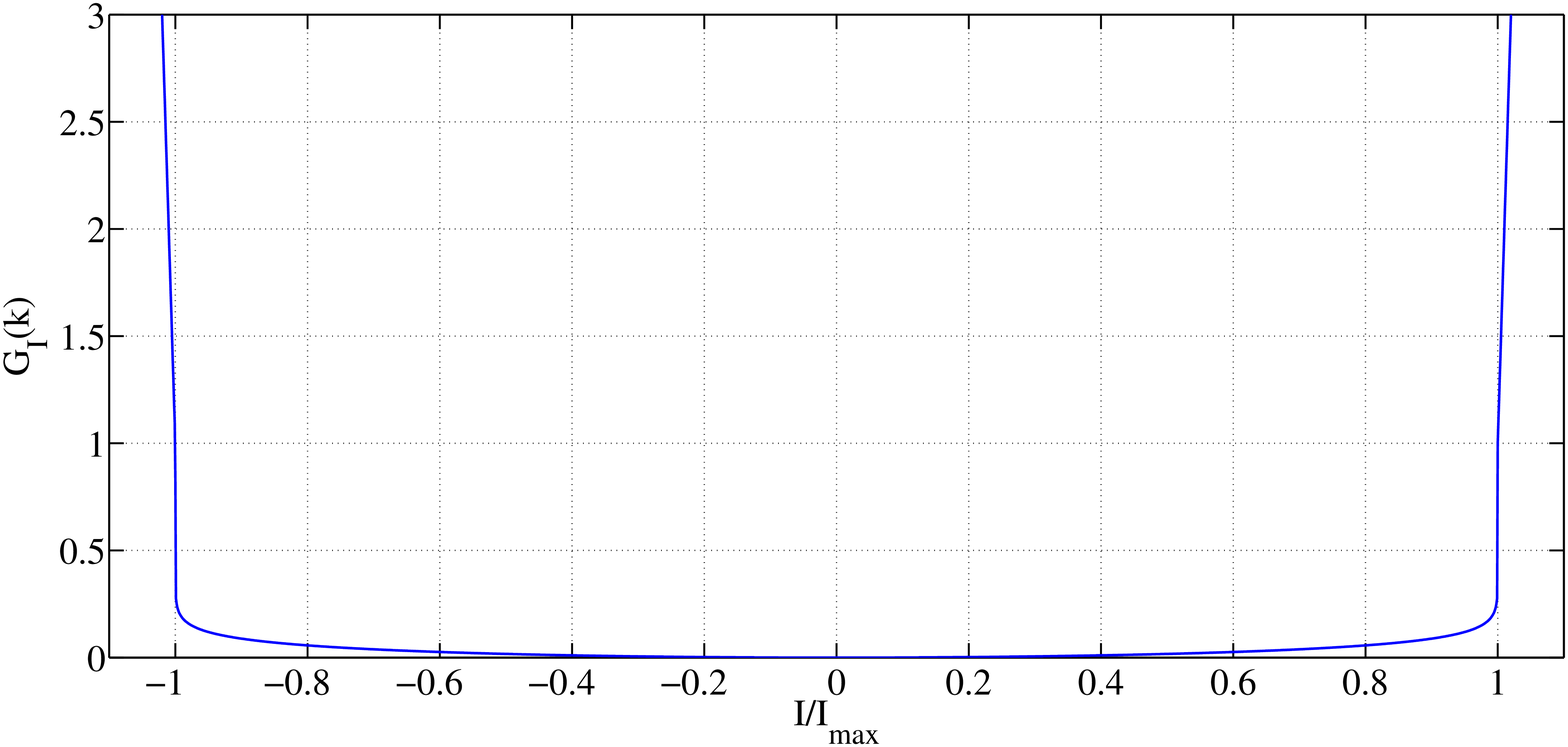} \\
(b)\\
\end{tabular}
\caption{Penalty functions: (a) $G_V(\cdot)$, (b) $G_I(\cdot)$.}
\label{fig::penalty_functions}
\end{figure}

The function $\Gamma(\mathbf{k})$ is defined as follows:
\begin{equation}
\Gamma(\mathbf{k})= (1-\beta) \Gamma_I(\mathbf{k}) + \beta \Gamma_V(\mathbf{k}),
\label{eq::beta}
\end{equation}
in which $\beta\in[0, 1]$ is a parameter used to adjust the relative weight of the violation of current constraints $\Gamma_I(\mathbf{k})$ with respect to the term related to voltages violation $\Gamma_V(\mathbf{k})$. In order to make the constraint violation value $\Gamma(\mathbf{k})$ of the same order as the cost function value $J(\mathbf{k})$, the functions $\Gamma_I(\mathbf{k})$ and $\Gamma_V(\mathbf{k})$ are defined as follows:
\begin{equation}
\label{eq::gamma}
\begin{aligned}
\Gamma_V(\mathbf{k})= & \max_{i\in\{0,..., N\}} \left\lbrace G_V\left( V_i(\mathbf{k})/V^{\mathrm{nom}}_i \right)  \right\rbrace, \\
\Gamma_I(\mathbf{k})= & \max_{j\in\{0,..., R\}} \left\lbrace G_I\left( I_j(\mathbf{k})/I^{\mathrm{max}}_j \right)  \right\rbrace, 
\end{aligned}
\end{equation}
where $V^{\mathrm{nom}}_i$ indicates the voltage nominal value on the $i$-th node.
The penalty functions used in (\ref{eq::gamma}), that is, $G_V(\cdot)$ and $G_I(\cdot)$ are graphically shown in Figure \ref{fig::penalty_functions}.
Further details of the optimization procedure can be found in \cite{Storti_2013}.
\subsection{Admissible Network Configurations}
\label{sec::AdmNetConf}

Consider a general SG consisting of several Medium Voltage (MV) feeders, some High Voltage (HV) substations, some DGs, and several loads. In order to perform the power loss minimization through DFR, we decided to represent the SG as a non-oriented graph $\mathcal{G} \langle N, E \rangle $, in which $N$ and $E$ are the nodes and the edges of the real network, respectively.
We introduce $\hat{\mathcal{G}} \langle \hat{N}, \hat{E} \rangle$, the \textit{reduced graph} of the network, to properly describe all possible system reconfigurations satisfying the topology constraints.
The reduced graph of the network $\hat{\mathcal{G}} \langle \hat{N}, \hat{E} \rangle$ does not contain all the information of the original network graph $\mathcal{G} \langle N, E \rangle $, because for our purposes we only need information about the connections of different portions of the network and not their detailed internal structure.
As described in \cite{Storti_2014}, $\mathcal{G} \langle N, E \rangle $ is mapped into $\hat{\mathcal{G}} \langle \hat{N}, \hat{E} \rangle$ through two main steps:
\begin{itemize}
\item 
The nodes $\hat{N}$ of the reduced graph are used to model two different types of original nodes $N$. The first one represents nodes at 150kV providing the energy balance of the active and reactive power in the SG. In the following sections we will refer to it as HV node. The second one can represent a single MV real substation connected to loads and DGs, or a set of MV substations, powered by only a single HV substation (virtual MV). In both cases, we call this kind of nodes as MV nodes.
\item 
Edges $\hat{E}$ of the reduced graph are used to model the topology reconfiguration. The series of two switches, installed between two consecutive MV substations, are mapped into a single edge (virtual breaker) of the reduced graph $\hat{\mathcal{G}}$. Each edge is associated with a label representing its state, i.e., close or open.
\end{itemize}

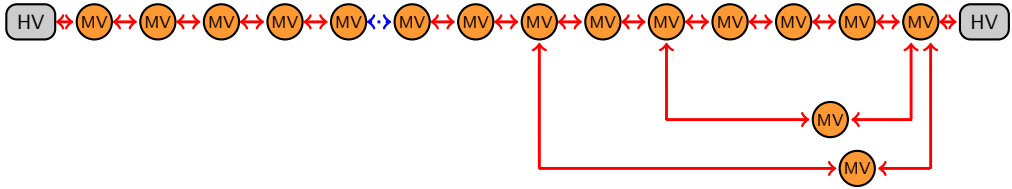
\begin{figure}[h!tbph]
\centering
\begin{tikzpicture}[<->, shorten >=1pt,auto,
    scale = 0.65, transform shape, thick,
    every node/.style = {draw, circle, minimum size = 7.2mm},
    grow = down,  
    level 1/.style = {sibling distance=2cm},
    level 2/.style = {sibling distance=2cm}, 
    level 3/.style = {sibling distance=2cm}, 
    level distance = 1.5cm,
    node distance=1.3cm,   
  ]
  \node[fill = gray!40, shape = rectangle, rounded corners,
    minimum width = 1cm, font = \sffamily] (Start) {HV};
  
  \node [head] (A) [right of = Start] {}; 
  \node [head] (B) [right of = A] {};
  \node [head] (C) [right of = B] {};
  \node [head] (D) [right of = C] {};
  \node [head] (E) [right of = D] {};
  \node [head] (F) [right of = E] {};
  \node [head] (G) [right of = F] {};
  \node [head] (H) [right of = G] {};
  \node [head] (I) [right of = H] {};
  \node [head] (L) [right of = I] {};
  \node [head] (M) [right of = L] {};
  \node [head] (N) [right of = M] {};
  \node [head] (O) [right of = N] {};
  \node [head] (P) [right of = O] {};
  
  \node[fill = gray!40, shape = rectangle, rounded corners,
    minimum width = 1cm, font = \sffamily, right of = P] (End) {HV};

  \path [red, line width=1.1pt] (Start.east) edge (A.west);
  \path [red, line width=1.1pt] (A.east) edge (B.west);
  \path [red, line width=1.1pt] (B.east) edge (C.west);
  \path [red, line width=1.1pt] (C.east) edge (D.west);
  \path [red, line width=1.1pt] (D.east) edge (E.west);
  \path [dotted, blue, line width=1.1pt] (E.east) edge (F.west);
  \path [red, line width=1.1pt] (F.east) edge (G.west);
  \path [red, line width=1.1pt] (G.east) edge (H.west);
  \path [red, line width=1.1pt] (H.east) edge (I.west);
  \path [red, line width=1.1pt] (I.east) edge (L.west);
  \path [red, line width=1.1pt] (L.east) edge (M.west);
  \path [red, line width=1.1pt] (M.east) edge (N.west);
  \path [red, line width=1.1pt] (N.east) edge (O.west);
  \path [red, line width=1.1pt] (O.east) edge (P.west);
  \path [red, line width=1.1pt] (P.east) edge (End.west);
  
  \node [head, name = Q, yshift = -30 mm] at (O) {}; 
  \node [name = QQ, yshift = -30 mm, draw = none, minimum size = 0 mm] at (H) {};
  \path [->, red, line width=1.1pt]([yshift=-1.2ex]QQ.north) edge (H.south);
  \path [->, red, line width=1.1pt] ([xshift=-1.2ex]QQ.east) edge (Q.west);
  \node [name = QQQ, yshift = -30 mm, draw = none, minimum size = 0 mm] at (P) {};
  \path [->, red, line width=1.1pt]([yshift=-1.2ex, xshift = 2mm]QQQ.north) edge ([xshift = 2mm]P.south);
  \path [->, red, line width=1.1pt] ([xshift=2.2ex]QQQ.west) edge (Q.east);
  
  \node [head, name = R, yshift = -20 mm, xshift = 7.5 mm] at (N) {}; 
  \node [name = RR, yshift = -20 mm, draw = none, minimum size = 0 mm] at (L) {}; 
  \path [->, red, line width=1.1pt]([yshift=-1.2ex]RR.north) edge (L.south);
  \path [->, red, line width=1.1pt] ([xshift=-1.2ex]RR.east) edge (R.west);
  \node [name = RRR, yshift = -20 mm, xshift = 5mm, draw = none, minimum size = 0 mm] at (P) {};
  \path [->, red, line width=1.1pt]([xshift=-7mm, yshift = -2mm]RRR.north) edge ([xshift = -2mm]P.south);
  \path [->, red, line width=1.1pt] ([xshift=-4.7mm]RRR.west) edge (R.east);
  
\end{tikzpicture}
\caption{Example of a graph $\hat{\mathcal{G}} \langle \hat{N}, \hat{E} \rangle$ of the simplified network. Yellow circles, gray rectangles, dashed blue arrows, and solid red arrows represent MV nodes, HV nodes, open status edges, and close status edges, respectively.}
\label{fig::graph}
\end{figure}

Figure \ref{fig::graph} shows an example of the representation of the SG through the reduced graph $\hat{\mathcal{G}} \langle \hat{N}, \hat{E} \rangle$.

Using the above notation we can introduce the following definitions:
\begin{thm}[\textbf{Radial Topology Constraint}]
A network topology satisfies the Radial Topology Constraint iff each MV substation is fed by only one HV substation via only one path.
\end{thm} 
\begin{thm}[\textbf{Admissible Configuration}]
\label{def::admConf}
A reduced graph $\hat{\mathcal{G}} \langle \hat{N}, \hat{E} \rangle$ satisfying the \textit{radial topology constraint} is said to be an admissible configuration of the network.
\end{thm}

The graph representation $\hat{\mathcal{G}} \langle \hat{N}, \hat{E} \rangle$ is used to execute an algorithm that performs an exhaustive search of all admissible configurations of the network. The details of the automatic procedure are described in Ref. \cite{Storti_2014}.
The output of such a procedure is a list of binary strings (encoding the admissible configurations) having length equals to the number of edges $ \hat{E}$ of the reduced network. Each bit represents the state of the corresponding edge (virtual breaker).
The network topology is specified through a label associated with the string of bits, spanning the rows of the list of all admissible configurations. Because of the nominal nature of the parameter specifying the network topology, the objective function becomes non-differentiable and the respective DFR optimization problem is very challenging.
For this reason, in this paper we use a heuristic method \cite{caschera_2014} based on the Hamming distance between network configurations, to improve the smoothness of the objective function with respect to the variation of the nominal parameter representing the topological configuration. 
This technique allows to treat during the optimization process the nominal parameter describing the configuration as an ordinal parameter.
The reader is referred to Ref. \cite{caschera_2014} for the details.

\section{The ACEA SG Pilot Network}
\label{sec:acea_opt}

In this work, we consider a portion of the Italian electric distribution network managed by ACEA Distribuzione S.p.A., located in the west area of Rome. The main ACEA goal is the overall improvement of the (i) quality of service related to the continuity of electricity distribution, (ii) capacity of the network, and (iii) prize of electricity offered to the users.
\subsection{Network Specifications}
\label{sec:Acea_SG}
The main specifics of the ACEA network are listed below:
\begin{itemize}
\item $N. 6$ Medium Voltage (MV) feeders ($n. 5$ at $20~kV$ and $n. 1$ at $8.4~kV$);
\item $N. 2$ High Voltage (HV) Substations;
\item $N. 76$ MV Substations ($n. 29$ at $20~kV$ and $n. 47$ at $8.4~kV$);
\item about $70~km$ of MV lines ($31~km$ of underground wires and $38~km$ of air lines);
\item about $1200$ Low Voltage (LV) user loads;
\item $N. 5$ DGs ($n. 5$ generator sets $n. 1$ photovoltaic generator);
\item $N. 106$ three-phase breakers;
\item $N. 1$ TVR (Thyristor Voltage Regulator).
\end{itemize}
In each HV substation there is a transformer that converts the voltage from 150 kV at the primary winding to 20 kV at the secondary winding (HV/MV transformer). 
The cables, the photovoltaic plant, the MV substations, and the TVR are located in the MV portion of the network, whereas the user loads and the five generator sets are located in the LV portion of the network.
The TVR is a series voltage compensation device. It performs a bi-directional voltage regulation that maintains the system voltage within specified ranges. 
%
%
The bi-directional relation between the input and the output voltage is defined as follows:
\begin{equation}
V_{\mathrm{out}} =  V_{\mathrm{in}} + N_{\mathrm{tap}} \Delta V, \quad N_{\mathrm{tap}} \in\{0,\pm 1,\pm 2,\pm 3\},
\label{eq::T}
\end{equation}
where the values of $V_{\mathrm{in}}$ and $V_{\mathrm{out}}$ are expressed in kV and the $\Delta V$ is 0.1~kV. The voltage rated value of $V_{\mathrm{in}}$ is 8.4~kV. 
Each MV substation is equipped with 2 breakers (switches) that allow to connect the substation with the electrical network in different ways. By changing the status of these switches, it is possible to modify the topology of the network.
The power quality is a very important issue in an electrical network, which in turn determines the quality of electrical power provided to consumer devices.
The correct setting of the electrical limits allows operating electrical systems in a safe way without significant loss of performance. In order to protect the customers, the Authority for Energy and Gas \footnote{http://www.autorita.energia.it/it/inglese/index.htm} has imposed constraints on voltage and current to power delivery companies:
\begin{itemize}
\item the instantaneous voltage of all the nodes of the network must be comprised in a range of $\pm~10~\%$ of nominal voltage;
\item the instantaneous current of all the branches of the network must be lower than a threshold.
\end{itemize}
These constraints are taken into consideration in the definition of penalty functions showed in Figure \ref{fig::penalty_functions}. If the voltage or current, measured at some nodes and branches of the network, exceed the admissible range, the value of the penalty function increases dramatically.
\subsection{Power Loss Optimization Problem Customization for the ACEA Network}
\label{sec::Opt_Prob_ACEA}

Using the network specifications described in Section \ref{sec:Acea_SG}, we can customize the optimization procedure introduced in Section \ref{sec::opt_procedure}.
In particular, we can control the reactive power of the five generator sets through their phase parameter, $\phi$.
On the other hand, it is not possible to control the reactive power of the photovoltaic generator. 
Moreover, it is possible to chose the $N_{\mathrm{tap}}$ value of the TVR and the topological configuration of the network selecting it from the set of admissible ones, previously determined.
The phases of the five generator sets, $\phi_1,\phi_2,\phi_3,\phi_4,\phi_5$, will be spanned in a real-valued range specified by the capability functions of the corresponding generator sets.
The tap $N_{\mathrm{tap}}$ of the TVR will be spanned in the discrete (normed) range defined in \eqref{eq::T}.
Finally, according to the list of admissible configurations computed using the procedure described in Section \ref{sec::AdmNetConf}, the network topology is specified by the index, $N_{\mathrm{conf}}$, spanning the rows of such a list.
In particular, in the SG under consideration, the total number of admissible configurations is 390.

Summarizing, a candidate solution vector of the optimization problem reads as $\mathbf{k} = [ \phi_1,\phi_2,\phi_3,\phi_4,\phi_5, N_{\mathrm{tap}}, N_{\mathrm{conf}} ]$ and technically belongs to the set $A$ defined by the following ranges,
\begin{eqnarray}
\label{eq::A}
-0.2 \leq  & \phi_1,\phi_2 & \leq 0.45 \\  \nonumber                        
-0.2 \leq  & \phi_3        & \leq 0.55 \\  \nonumber
0.0 \leq  & \phi_4        & \leq 0.64 \\  \nonumber
-0.32 \leq  & \phi_5        & \leq 0.45 \\  \nonumber
-3 \leq  & N_{\mathrm{tap}}       & \leq 3 \\ \nonumber
 N_{\mathrm{conf}} \in & \{1,..., 390\},                       
\end{eqnarray}
where $\{1,..., 390\}$ is the set of indexes of all admissible configurations. We remind that, due to the heuristic ordering mentioned in Section \ref{sec::AdmNetConf}, we treat the nominal parameter $N_{\mathrm{conf}}$ as ordinal during the optimization procedure.
Moreover, according with the Authority requirements, in order to be valid, a candidate solution $\mathbf{k}$ must satisfy the constraints on voltages and currents defined below:
\begin{equation}
\begin{aligned}
\label{eq:vinc}
B &= \big\{ \mathbf{k} \subset A : 0.9 V^{\mathrm{nom}}_i \leq V_i(\mathbf{k})\leq 1.1 V^{\mathrm{nom}}_i , i=1,...,N   \big\} \\
C &=  \left\{ \mathbf{k} \subset A  : |I_j(\mathbf{k})|\leq I^{\mathrm{max}}_j, j=1,...,R \right\},
\end{aligned}
\end{equation}
in which $N$ and $R$ represent the total number of nodes and branches of the real network, respectively, whereas $V^{\mathrm{nom}}_i$ and $I^{\mathrm{max}}_j$ are the nominal value of the voltage of the $i$-th node and the maximum current allowed in the $j$-th wire, respectively.

\section{Analysis of Admissible Configurations}
\label{sec:dis_FF}

According to Def. \ref{def::admConf}, a reduced graph $\hat{\mathcal{G}} \langle \hat{N}, \hat{E} \rangle$ satisfying the \textit{Radial Topology Constraint} is an \textit{admissible configuration} for the reference SG.
In Section \ref{sec::opt_procedure} we introduced the problem of minimization of power loss $J(\mathbf{k})$ defined in the domain $E$ in terms of the minimization of the convex function $F(\mathbf{k})$ in $A$.
It is worth noting that, in order to make the optimization problem well posed, both terms of \eqref{eq::alpha} must be normalized in the same range (e.g., $[0, 1]$).
During several preliminary tests, we noticed that for a certain number of runs of the electrical network simulation, the objective function, $F(\mathbf{k})$, assumes values greater than unity, hence violating such a requirement.
In these situations the GA seems to have difficulties in minimizing the power losses, $J(\mathbf{k})$, during the optimization process.
According to a preliminary analysis of this phenomenon, this behavior seems to be related only to the actual value of the topological parameter $N_{\mathrm{conf}}$.
In fact, for a certain selection of the $N_{\mathrm{conf}}$ parameter by the GA, the constraint term $\Gamma(\mathbf{k})$ seems to increase dramatically compared to the $J(\mathbf{k})$ term, causing the power losses minimization procedure to fail.
To better understand this aspect, in the following section we perform a numerical analysis of admissible configurations of the ACEA SG in terms of the violation of electrical constraints and we introduce the concept of \textit{Constraint Compliant Configurations}.
%
%
\subsection{Constraint Compliant Configurations}
First we introduce the concept of Constraint Compliant Configurations (CCC) through the following definition.
\begin{thm}[\textbf{Constraint Compliant Configuration}]
A reduced graph $\hat{\mathcal{G}} \langle \hat{N}, \hat{E} \rangle$ whose nodes satisfy the voltage constraints and edges satisfy the current constraint (\ref{eq:vinc}) is said to be a CCC.
\end{thm} 

We performed several experiments to numerically show that in the SG under analysis there exist admissible configurations that inherently violate constraints on voltage at some nodes and/or on current at some branches. This means that, independently of the choice of the parameters $ \phi_1$, $\phi_2$, $\phi_3$, $\phi_4$, $\phi_5$,  $N_{\mathrm{tap}}$ of the network, constraints in \eqref{eq:vinc} are violated.
Roughly speaking, considering an admissible configuration represented by its reduced graph $\hat{\mathcal{G}}_i \langle \hat{N}, \hat{E} \rangle$, and the corresponding configuration parameter $N_{\mathrm{conf}}=i$, the value of the objective function $F(\mathbf{k})$ associated with it remains of the same order of magnitude as the remaining parameters change.
To support this assertion, we analyze all admissible configurations. In particular, once the configuration parameter $N_{\mathrm{conf}}$ is set, we perform a random sampling of the subset $A$ defined in \eqref{eq::A} along all other dimensions. More precisely, we randomly choose 2000 points of the subset $A\setminus\mathbb{X}$.

\begin{figure}[htbph]
\centering
\begin{tabular}{cc}
\includegraphics[viewport=0 0 374 287,scale=0.5]{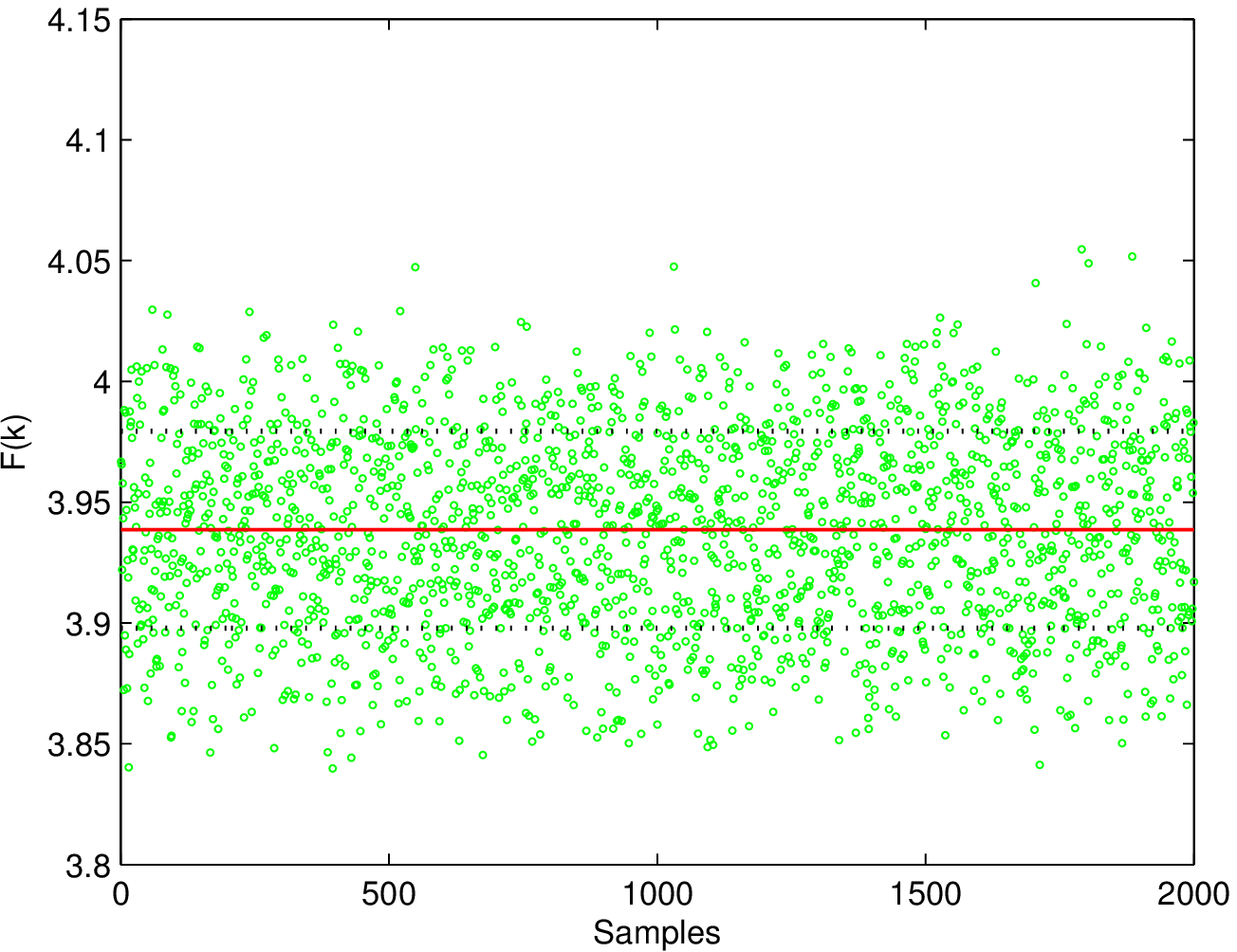} &
\includegraphics[viewport=0 0 385 287,scale=0.5]{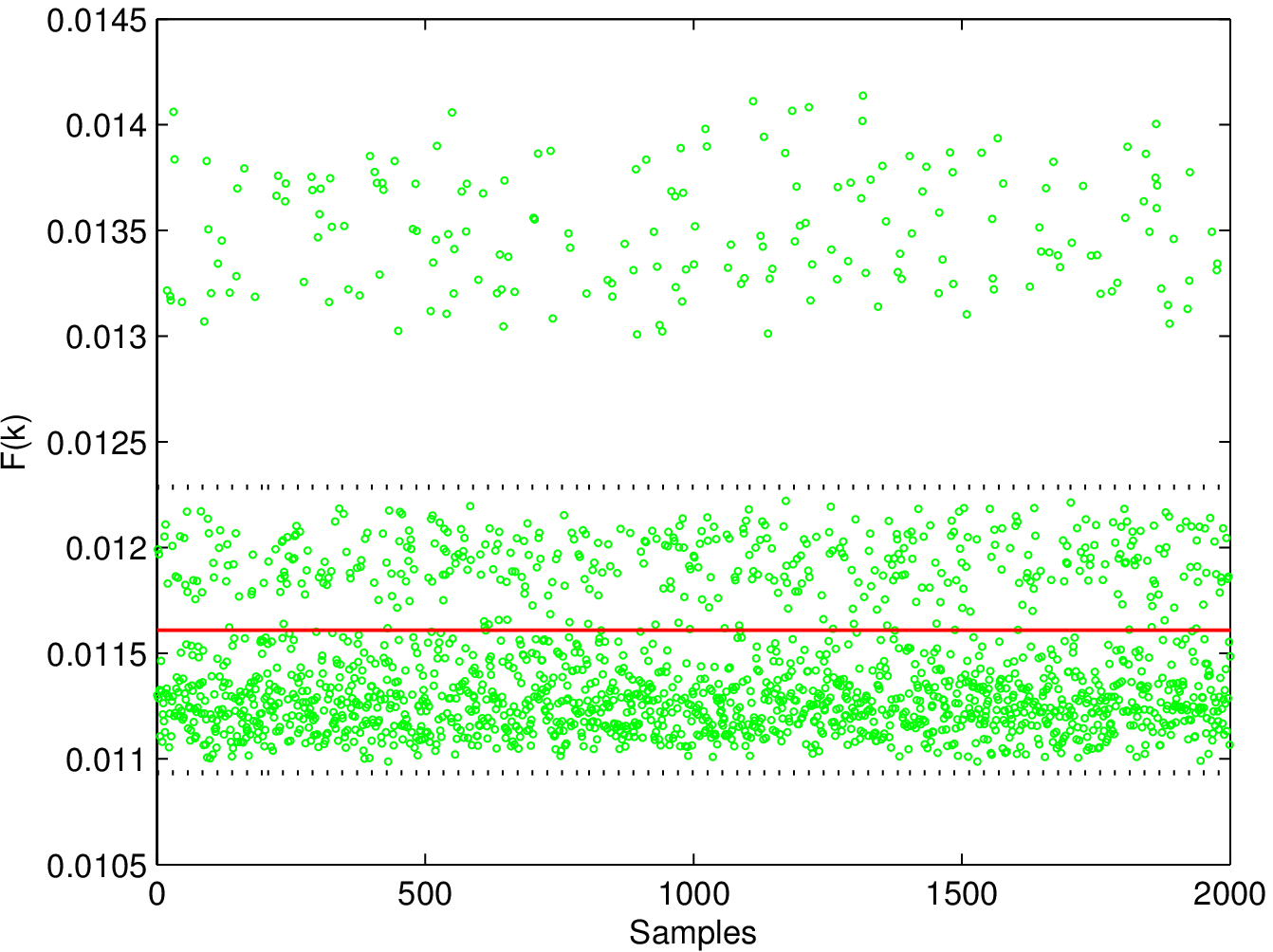} \\
(a)&(b)\\
\end{tabular}
\caption{Objective function $F(\mathbf{k})$ computed at 2000 randomly selected points, maintaining constant the parameter (a) $N_{\mathrm{conf}} = 32$, (b) $N_{\mathrm{conf}} = 81$. The mean value is highlighted in red, while the standard deviation in dotted black lines.}
\label{fig:campionamentoAlto_Basso}
\end{figure}

The performed analysis highlights two different behaviors. Here, we report the results for two admissible configurations associated with the two different behaviors.
The results obtained for the two configurations labeled $32$ and $81$ are shown in Figure \ref{fig:campionamentoAlto_Basso} parts (a) and (b), respectively.

It is worth noting in Figure \ref{fig:campionamentoAlto_Basso} part (b) that the values of the objective function $F(\mathbf{k})$ are distributed in bands due to the influence of the parameter $N_{\mathrm{tap}}$.
Comparing part (a) and (b) of Figure \ref{fig:campionamentoAlto_Basso}, it can be observed that the mean value (shown as a red line in the figure) of the objective function $F(\mathbf{k})$ is much higher (lower) than unity in the first (second) configuration. 
This fact shows that, if the parameter $N_{\mathrm{conf}}=32$ is selected, the GA will be probably unable to make $F(\mathbf{k})$ lower then unity. In fact, the order of magnitude of the objective function value depends only on the parameter $N_{\mathrm{conf}}$ and it does not change with the other parameters.
Moreover, by computing the ratio $\eta= \sigma / \mu$ between the standard deviation and the mean value for the 32-\textit{th} and 81-\textit{th} configurations we find $\eta = 0.0104 $ and $\eta = 0.0582 $, respectively. This means that in configuration number 32 the variation of the control parameters have a lower influence in the minimization of $F(\mathbf{k})$ with respect to configuration number 81.

\begin{figure}[!thbph]
\centering
\begin{tabular}{cc}
\hspace{-0.5cm}
\includegraphics[viewport=0 0 385 287,scale=.5]{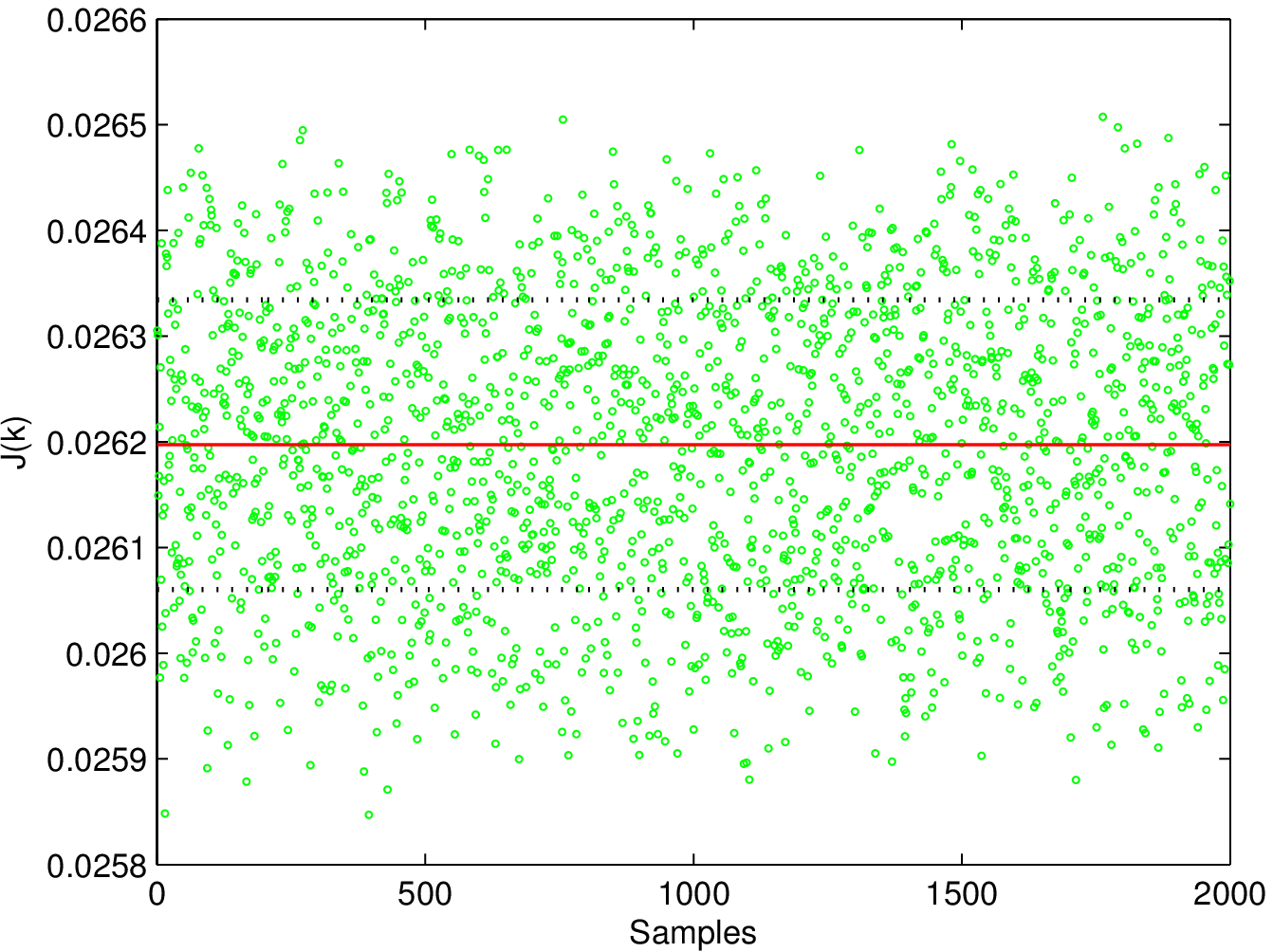} &
\includegraphics[viewport=0 0 375 287,scale=.5]{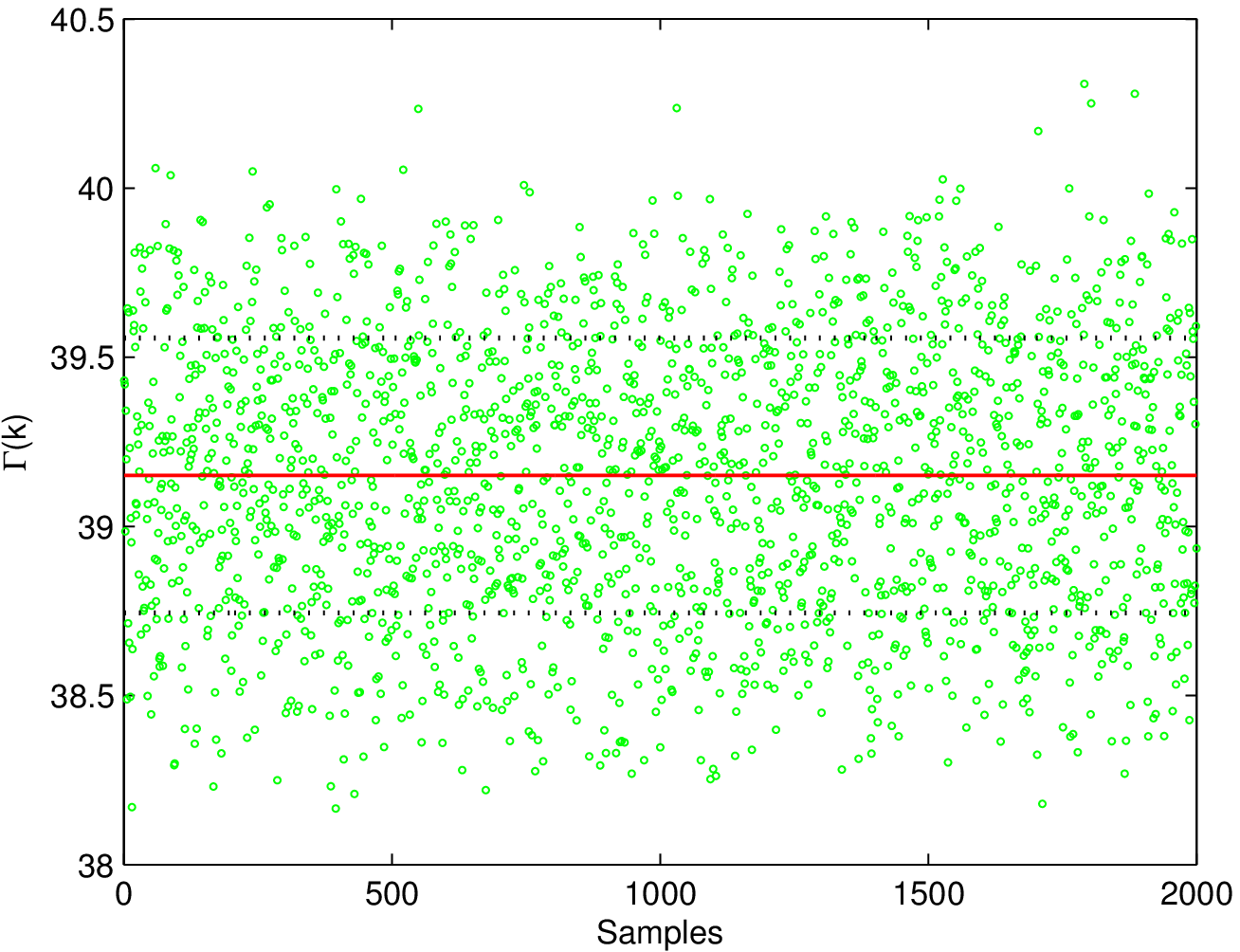}\\
(a)&(b)\\
\end{tabular}
\caption{Random sampling of the subset $A$, maintaining constant the parameter $N_{\mathrm{conf}}=32$. (a) Power Loss Term $J(\mathbf{k})$, (b) Constraints Term $\Gamma(\mathbf{k})$.}
\label{fig::J_gamma_conf32}
\end{figure}

To better understand this fact, we analyze the different components of \eqref{eq::alpha} for the 32-\textit{th} network configuration.
The value of the two terms, $J(\mathbf{k})$ and $\Gamma(\mathbf{k})$, for all the samples and the respective mean values and standard deviations are shown in Figure \ref{fig::J_gamma_conf32}. Remember from \eqref{eq::J} and \eqref{eq::beta} that the term $J(\mathbf{k})$ is responsible for the minimization of the power loss, while the term $\Gamma(\mathbf{k})$ guarantees that a configuration violating the constraints on voltages and/or currents is severely penalized by the optimization algorithm itself (insuring a very low fitness value).
From Figure \ref{fig::J_gamma_conf32} it is possible to observe that the increase in the objective function is due to the term $\Gamma(\mathbf{k})$, meaning that for this configuration the constraints \eqref{eq:vinc} are violated, and all parameters (except for $N_{\mathrm{conf}}$) fail to bring the network in a safe condition.
We can conclude that the 81-\textit{th} configuration is CCC, while the 32-\textit{th} is not. 

\subsection{Influence of NCCCs on the Objective Function}
As a consequence of what we have empirically shown in the previous section, during the optimization procedure candidates that are admissible configurations but not CCC will never be solution of the optimization problem \eqref{eq::alpha}.
The presence of such configurations in the solution space negatively affects the behavior of the optimization procedure, as we will discuss in detail in Section \ref{sec::ExpRes}.

Here we give a practical demonstration of the negative influence of those configurations that are not constraint compliant (shortened as NCCC) in the objective function \eqref{eq::alpha}.
We know that in the convex combination \eqref{eq::alpha}, the real parameter $\alpha$ is used to adjust the relative weight of the two terms of the expression. Moreover, in order for the optimization problem to be well posed, both $J(\mathbf{k})$ and $\Gamma(\mathbf{k})$ must be normalized in the same range (in this case $[0, 1]$).
However, in the previous section we have shown that for all admissible configurations, the terms $J(\mathbf{k})$ and $\Gamma(\mathbf{k})$ are not necessarily normalized in $[0, 1]$.
In particular, by defining $J_{\mathrm{MAX}}(\mathbf{k})$ as the maximum value that the power loss can assume and $\Gamma_{\mathrm{MAX}}(\mathbf{k})$ as the maximum value for $\Gamma(\mathbf{k})$, from \eqref{eq::alpha}, we can derive the following expression:
\begin{equation}
F(\mathbf{k}) = \alpha J_N(\mathbf{k}) \cdot J_{\mathrm{MAX}}(\mathbf{k}) + (1-\alpha) \Gamma_N(\mathbf{k}) \cdot \Gamma_{\mathrm{MAX}}(\mathbf{k}),
\label{eq::F_normaliz}
\end{equation}
in which $J_N(\mathbf{k})$ and $\Gamma_{N}(\mathbf{k})$ are the normalized values in $[0, 1]$ of the corresponding functions -- normalization is implemented by dividing for the maximum value.

In the following, we derive a value for the weighting parameter $\alpha$, denoted as $\alpha_{\mathrm{eq}}$, such that the convex combination in \eqref{eq::F_normaliz} is safely mapped to the unit interval.
Let us define $\alpha_{\mathrm{eq}}$ as the equivalent (i.e., transformed) $\alpha$ coefficient that should be used in $F(\mathbf{k})$ if its terms were normalized.
Using simple algebra (details of the calculations are provided in \ref{sec:appendixA}) we can derive the analytical expression for $\alpha_{\mathrm{eq}}$:
\begin{equation}
\alpha_{\mathrm{eq}} = \frac{\alpha J_{\mathrm{MAX}}(\mathbf{k})}{\alpha J_{\mathrm{MAX}}(\mathbf{k}) + (1-\alpha)\Gamma_{\mathrm{MAX}}(\mathbf{k})}.
\label{eq::alpha_eff}
\end{equation}

As an example, for the 32-\textit{th} configuration considered in the previous section, the maximum value of the constraint term is $\Gamma_{\mathrm{MAX}}(\mathbf{k}) = 40.38$ and for the power loss is $J_{\mathrm{MAX}}(\mathbf{k}) = 0.0265$.
Therefore, assuming that in principle we want to give more weight to the minimization of the power loss, i.e., by setting $\alpha = 0.9$, during the optimization using  \eqref{eq::alpha_eff} the effective weight is set to $\alpha_{\mathrm{eq}} \simeq 0.0059$.
This means that for all individuals in the GA population with the $N_{\mathrm{conf}}$ parameter set to indexes corresponding to NCCCs, the algorithm is almost entirely devoted to reducing the value of the constraint term instead of the power loss term.
However, in the previous sections we have empirically demonstrated that, for such configurations, it is practically impossible to bring the network in a safe condition, regardless of the setting of all other parameters.
This fact has a negative impact on the entire optimization process, as we will further discuss in the following section.

\section{Simulation Results}
\label{sec::ExpRes}

In this section, we first compare the GA performances when it is used to solve the optimization problem presented in Section \ref{sec::opt_procedure} for the electrical network presented in Section \ref{sec:Acea_SG}, whether in the admissible set $A$ are included only CCCs or all admissible configurations (both NCCCs and CCCs).
Successively, we also give an electrical interpretation of the obtained results.

\subsection{Optimization Results}
Among the 390 admissible configurations described in Section \ref{sec::Opt_Prob_ACEA}, we first manually extract all configurations that are CCCs. The result is that 372 are uniquely classified: 151 are CCCs, while the remaining 221 are NCCCs.
We want to compare the performances of the GA in solving the optimization problem \ref{sec::Opt_Prob_ACEA} when in the solution space are present or not the NCCCs.
Let us refer to the case with all admissible network configurations as ``Experiment 1'', while with ``Experiment 2'' we refer to the case considering CCCs only.
In both experiments, we employ the ordering criteria for the network configurations described in Ref. \cite{caschera_2014}.
We use a simulation model of the ACEA network \cite{Storti_2013} realized using MATLAB and Simulink, together with the GA implemented as described in Ref. \cite{deep2009real}.
To perform all experiments, we consider as input of the network model the power profile of distributed generators and loads registered at 1:00PM (for one hour) on January 1\textit{st}.

In the following, we provide some relevant details about the setting of the GA.
The number of individuals in the GA population is set to $20$; the elite individuals are $2$, i.e., only $2$ individuals in the current generation are guaranteed to survive in the next generation; the crossover fraction parameter is $0.8$; the mutation operator is applied to the remaining individuals with rate $0.1$. Furthermore, the $\alpha$ and $\beta$ coefficients used in expressions (\ref{eq::alpha}) and (\ref{eq::beta}) are set to $0.9$ and $0.2$, respectively.
The maximum number of iterations before the algorithm halts is $100$, but the GA might stop if the relative change of the fitness value over $50$ iterations is less than or equal to $10^{-9}$.
We execute the GA ten different times with different random initialization seeds; the $j$-th execution considers an equivalent initial population $P_j$ for both series of simulations (i.e. only CCCs or all admissible configurations in the solution space).
However, for simulations that consider CCCs only, the individuals of $P_j$, whose $7$-th gene (corresponding to the $N_{\mathrm{conf}}$ variable) specifies a NCCC are replaced by randomly generated individuals, whose $7$-th gene is forced to code for a CCC.

Results of the simulations are shown in Table \ref{tab::performanceUltimoGA}. The table reports the mean value and standard deviation of the number of generations ($\#\mathrm{gen}$) required for convergence, the fitness value percentage reduction ($\Delta F$), and the reduction of power loss in the network ($\Delta P_{\mathrm{loss}}$) for both experiment settings.
More precisely, the last two indicators compare the fitness value and actual power loss at optimal solution with respect to fitness value and power loss of the best individual in the initial population.
\begin{table}[htbph]
\centering
\caption{Mean and standard deviation for number of generations (\# $\mathrm{gen}$) required by the GA converge, reduction of the fitness value expressed in percentage ($\Delta F$), and reduction of the power loss ($\Delta P_{\mathrm{loss}}$) in Experiment 1 (all configurations) and Experiment 2 (CCCs only).}
\begin{tabular}{c|c|c|}
\cline{2-3}
 & \textbf{Experiment 1} & \textbf{Experiment 2} \\
\hline
\multicolumn{1}{ |c|  }{\# $\mathrm{gen}$} & $73 \pm18.3763$ & $65\pm 11.6858$\\ \hline
\multicolumn{1}{ |c|  }{$\Delta F$ [\%]} & $0.0227\pm 0.0001$ & $0.0205 \pm 0.0010$\\ \hline
\multicolumn{1}{ |c|  }{$\Delta P_{\mathrm{loss}}$ $[W]$} & $1097\pm 6.5015$ & $1138\pm 9.6788$ \\ \hline
\end{tabular}
\label{tab::performanceUltimoGA}
\end{table}

By analyzing the results in Table \ref{tab::performanceUltimoGA}, we can observe that (i) there is no statistically significant difference for the average number of iterations required for convergence and (ii) the mean reduction of the achieved fitness function value in Experiment 2 is slightly less (but statistically significant) than the one observed for Experiment 1. This second fact would led us to assume an inferior mean reduction of the power loss in Experiment 2 as well.
However, the power loss reduction achieved in Experiment 2 is significantly higher than the one observed for Experiment 1 -- please note that differences are statistically significant and are evaluated with t-test, $p<0.0001$.
The reason behind this result can be found in the redefinition of the solution set considered for the optimization problem in Experiment 2, composed by CCCs only.
In fact, in the set of network configurations considered in Experiment 1, many candidate solutions systematically violate constraints on voltages and currents causing the problems described in the previous section. 
Such a behavior is even more accentuated by the fact that we have chosen a small value of the coefficient $\beta$, and configurations that do not satisfy the constraints cause particularly very strong violations of the electrical current constraints.     
At first, the herein reported overall power loss reduction might appear not very significant. However, we remind that all tests simulate only one hour of one specific day of the year. Projecting the achieved power loss reduction to the entire network over a longer period of time might led to important improvements of the operating condition of the SG managed by ACEA.

In conclusion, we demonstrated that removing from the set of admissible configurations those that are NCCCs causes a significant reduction of the power loss.
This fact is particularly important in order to adopt such an optimization approach in a real time control system, where desired solutions must be determined on a hourly basis, usually relying on limited computational resources.

\subsection{Electrical Interpretation of the Representative Network Configurations}
\label{sec::clustering_analysis}
As observed in the previous sections, among the admissible configurations it is possible to recognize a subset where current or voltage constraints are systematically violated (NCCCs), regardless of the DGs setting.

In order to provide a meaningful electrical interpretation for this fact, a single representative network configuration for both sets of CCCs and NCCCs is selected.
Let $\mathcal{S}$ be a set of $n$ graphs (network configurations in our case).
A natural candidate to represent $\mathcal{S}$ is the graph $\hat{\mathcal{G}}^*$ that minimizes the sum of distances (MinSOD) \cite{delvescovo+livi+rizzi+frattalemascioli2011}, which is determined by the following expression:
\begin{equation}
\label{eq:minsod}
\hat{\mathcal{G}}^* = \argmin_{\hat{\mathcal{G}}_j\in\mathcal{S}} \sum_{i=1}^n d(\hat{\mathcal{G}}_j, \hat{\mathcal{G}}_i).
\end{equation}

In Eq. \ref{eq:minsod}, $d(\cdot, \cdot)$ is a dissimilarity measure between graphs \cite{gm_survey}, which in our case is implemented as the Hamming distance among the adjacency matrix representations of graphs -- we remind that we are considering Boolean graphs of the same order, i.e., graphs with the same number of nodes entirely described by the presence or absence of edges.
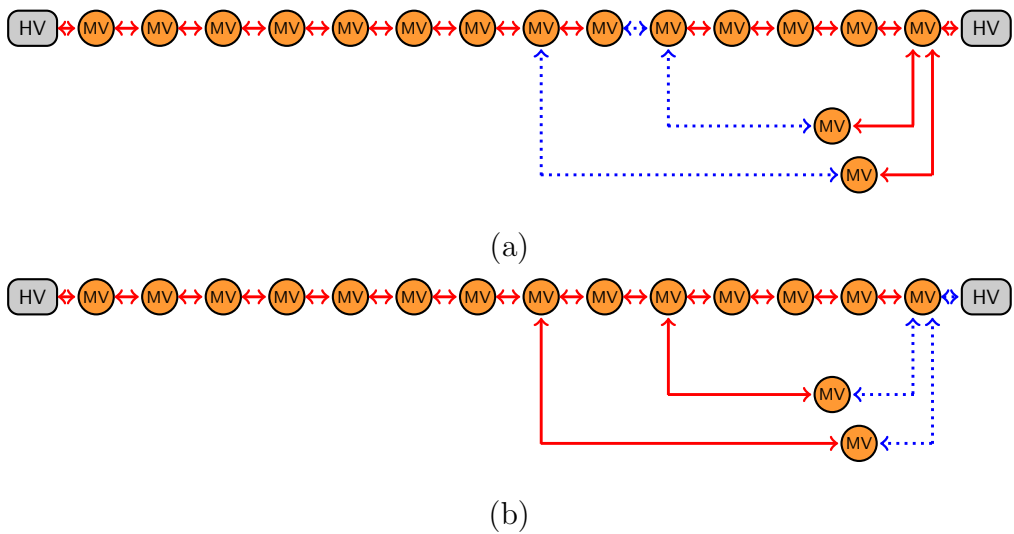
\begin{figure}[!htbph]
\centering
\begin{tabular}{c}
\begin{tikzpicture}[<->, shorten >=1pt,auto,
    scale = 0.65, transform shape, thick,
    every node/.style = {draw, circle, minimum size = 7.2mm},
    grow = down,  
    level 1/.style = {sibling distance=2cm},
    level 2/.style = {sibling distance=2cm}, 
    level 3/.style = {sibling distance=2cm}, 
    level distance = 1.5cm,
    node distance=1.3cm,   
  ]
  \node[fill = gray!40, shape = rectangle, rounded corners,
    minimum width = 1cm, font = \sffamily] (Start) {HV};
  
  \node [head] (A) [right of = Start] {}; 
  \node [head] (B) [right of = A] {};
  \node [head] (C) [right of = B] {};
  \node [head] (D) [right of = C] {};
  \node [head] (E) [right of = D] {};
  \node [head] (F) [right of = E] {};
  \node [head] (G) [right of = F] {};
  \node [head] (H) [right of = G] {};
  \node [head] (I) [right of = H] {};
  \node [head] (L) [right of = I] {};
  \node [head] (M) [right of = L] {};
  \node [head] (N) [right of = M] {};
  \node [head] (O) [right of = N] {};
  \node [head] (P) [right of = O] {};
  
  \node[fill = gray!40, shape = rectangle, rounded corners,
    minimum width = 1cm, font = \sffamily, right of = P] (End) {HV};

  \path [red, line width=1.1pt] (Start.east) edge (A.west);
  \path [red, line width=1.1pt] (A.east) edge (B.west);
  \path [red, line width=1.1pt] (B.east) edge (C.west);
  \path [red, line width=1.1pt] (C.east) edge (D.west);
  \path [red, line width=1.1pt] (D.east) edge (E.west);
  \path [red, line width=1.1pt] (E.east) edge (F.west);
  \path [red, line width=1.1pt] (F.east) edge (G.west);
  \path [red, line width=1.1pt] (G.east) edge (H.west);
  \path [red, line width=1.1pt] (H.east) edge (I.west);
  \path [dotted, blue, line width=1.1pt] (I.east) edge (L.west);
  \path [red, line width=1.1pt] (L.east) edge (M.west);
  \path [red, line width=1.1pt] (M.east) edge (N.west);
  \path [red, line width=1.1pt] (N.east) edge (O.west);
  \path [red, line width=1.1pt] (O.east) edge (P.west);
  \path [red, line width=1.1pt] (P.east) edge (End.west);
  
  \node [head, name = Q, yshift = -30 mm] at (O) {}; 
  \node [name = QQ, yshift = -30 mm, draw = none, minimum size = 0 mm] at (H) {};
  \path [->, dotted, blue, line width=1.1pt]([yshift=-1.2ex]QQ.north) edge (H.south);
  \path [->, dotted, blue, line width=1.1pt] ([xshift=-1.2ex]QQ.east) edge (Q.west);
  \node [name = QQQ, yshift = -30 mm, draw = none, minimum size = 0 mm] at (P) {};
  \path [->, red, line width=1.1pt]([yshift=-1.2ex, xshift = 2mm]QQQ.north) edge ([xshift = 2mm]P.south);
  \path [->, red, line width=1.1pt] ([xshift=2.2ex]QQQ.west) edge (Q.east);
  
  \node [head, name = R, yshift = -20 mm, xshift = 7.5 mm] at (N) {}; 
  \node [name = RR, yshift = -20 mm, draw = none, minimum size = 0 mm] at (L) {}; 
  \path [->, dotted, blue, line width=1.1pt]([yshift=-1.2ex]RR.north) edge (L.south);
  \path [->, dotted, blue, line width=1.1pt] ([xshift=-1.2ex]RR.east) edge (R.west);
  \node [name = RRR, yshift = -20 mm, xshift = 5mm, draw = none, minimum size = 0 mm] at (P) {};
  \path [->, red, line width=1.1pt]([xshift=-7mm, yshift = -2mm]RRR.north) edge ([xshift = -2mm]P.south);
  \path [->, red, line width=1.1pt] ([xshift=-4.7mm]RRR.west) edge (R.east);
  
\end{tikzpicture}
\\
(a)
\\
\begin{tikzpicture}[<->, shorten >=1pt,auto,
    scale = 0.65, transform shape, thick,
    every node/.style = {draw, circle, minimum size = 7.2mm},
    grow = down,  
    level 1/.style = {sibling distance=2cm},
    level 2/.style = {sibling distance=2cm}, 
    level 3/.style = {sibling distance=2cm}, 
    level distance = 1.5cm,
    node distance=1.3cm,   
  ]
  \node[fill = gray!40, shape = rectangle, rounded corners,
    minimum width = 1cm, font = \sffamily] (Start) {HV};
  
  \node [head] (A) [right of = Start] {}; 
  \node [head] (B) [right of = A] {};
  \node [head] (C) [right of = B] {};
  \node [head] (D) [right of = C] {};
  \node [head] (E) [right of = D] {};
  \node [head] (F) [right of = E] {};
  \node [head] (G) [right of = F] {};
  \node [head] (H) [right of = G] {};
  \node [head] (I) [right of = H] {};
  \node [head] (L) [right of = I] {};
  \node [head] (M) [right of = L] {};
  \node [head] (N) [right of = M] {};
  \node [head] (O) [right of = N] {};
  \node [head] (P) [right of = O] {};
  
  \node[fill = gray!40, shape = rectangle, rounded corners,
    minimum width = 1cm, font = \sffamily, right of = P] (End) {HV};

  \path [red, line width=1.1pt] (Start.east) edge (A.west);
  \path [red, line width=1.1pt] (A.east) edge (B.west);
  \path [red, line width=1.1pt] (B.east) edge (C.west);
  \path [red, line width=1.1pt] (C.east) edge (D.west);
  \path [red, line width=1.1pt] (D.east) edge (E.west);
  \path [red, line width=1.1pt] (E.east) edge (F.west);
  \path [red, line width=1.1pt] (F.east) edge (G.west);
  \path [red, line width=1.1pt] (G.east) edge (H.west);
  \path [red, line width=1.1pt] (H.east) edge (I.west);
  \path [red, line width=1.1pt] (I.east) edge (L.west);
  \path [red, line width=1.1pt] (L.east) edge (M.west);
  \path [red, line width=1.1pt] (M.east) edge (N.west);
  \path [red, line width=1.1pt] (N.east) edge (O.west);
  \path [red, line width=1.1pt] (O.east) edge (P.west);
  \path [dotted, blue, line width=1.1pt] (P.east) edge (End.west);
  
  \node [head, name = Q, yshift = -30 mm] at (O) {}; 
  \node [name = QQ, yshift = -30 mm, draw = none, minimum size = 0 mm] at (H) {};
  \path [->, red, line width=1.1pt]([yshift=-1.2ex]QQ.north) edge (H.south);
  \path [->, red, line width=1.1pt] ([xshift=-1.2ex]QQ.east) edge (Q.west);
  \node [name = QQQ, yshift = -30 mm, draw = none, minimum size = 0 mm] at (P) {};
  \path [->, dotted, blue, line width=1.1pt]([yshift=-1.2ex, xshift = 2mm]QQQ.north) edge ([xshift = 2mm]P.south);
  \path [->, dotted, blue, line width=1.1pt] ([xshift=2.2ex]QQQ.west) edge (Q.east);
  
  \node [head, name = R, yshift = -20 mm, xshift = 7.5 mm] at (N) {}; 
  \node [name = RR, yshift = -20 mm, draw = none, minimum size = 0 mm] at (L) {}; 
  \path [->, red, line width=1.1pt]([yshift=-1.2ex]RR.north) edge (L.south);
  \path [->, red, line width=1.1pt] ([xshift=-1.2ex]RR.east) edge (R.west);
  \node [name = RRR, yshift = -20 mm, xshift = 5mm, draw = none, minimum size = 0 mm] at (P) {};
  \path [->, dotted, blue, line width=1.1pt]([xshift=-7mm, yshift = -2mm]RRR.north) edge ([xshift = -2mm]P.south);
  \path [->, dotted, blue, line width=1.1pt] ([xshift=-4.7mm]RRR.west) edge (R.east);
  
\end{tikzpicture}
\\
(b)
\end{tabular}
\caption{MinSOD graphs for (a) CCC and (b) NCCC network configuration classes.}
\label{fig::minsod}
\end{figure}

We computed the MinSOD element for the CCC and NCCC classes separately. The MinSOD graphs are graphically represented in Figure \ref{fig::minsod}.
First of all, it can be noted that for the representative of NCCCs (\ref{fig::minsod} part (b)) all MV stations are fed by a single HV node, whereas the second available HV node is electrically isolated. This configuration results in a very long feeder. As it is well-known, long feeders can suffer of low-voltage problems in the nodes far from the feed bar (HV nodes) and can exhibit over-current issues close to the feed bar due to the large amount of user load connected to it.
For this reason, configurations containing very long feeders are very likely to be NCCCs, regardless of the DGs setting. 
Conversely, by analyzing the representative graph of CCCs (\ref{fig::minsod} part (a)) it is possible to note that user loads are distributed between the two available HV nodes, resulting in shorter and more equilibrated feeders having higher probability to fulfil constraints on voltage and current.

This observation suggests that, for each admissible network configuration considered in the optimization problem, the user load should be balanced among all the available HV nodes as much as possible, in order to reduce the typical length of the feeders and the corresponding amount of required electrical current at the feed bar.

\section{Conclusions}
\label{sec:conclusions}

In this paper we have presented an improvement of the control system first described in \cite{Storti_2013,Possemato_2013,Storti_2013_b,Storti_2014}.
We performed an analysis of admissible network configurations to identify undesirable configurations that may slow down the convergence speed of the optimization procedure. 
In particular, we have shown that there exist few admissible configurations for which it is not possible to avoid voltage/current constraint violation. 
We performed experiments on real data concerning one hour of power profile of distributed generators and loads for the SG located in the west area of Rome, realized by the company ACEA Distribuzione.
Results showed that, for the network under analysis, removing from the solution space of the optimization problem those configurations that are not constraint compliant leads to overall improvements in terms of the power loss reduction.
Successively, we used graph-based pattern analysis techniques to identify representative networks of both the constraint compliant and the undesirable configuration classes.
We then interpreted the results of such analysis from the electrical point of view. In particular, we noted that only those configurations with user loads suitably balanced among all the available HV nodes can be actual solutions of the considered optimization problem.
Accordingly, configurations having very long feeders might be a priori neglected from the search space, without causing loss of performance in the optimization of the system.
In future works, we intend to repeat such analysis for an extended portion of the network by considering also different time periods. Moreover, we intend to verify if it is possible, using pattern recognition techniques, to predict the type of configuration without simulating the entire network, improving thus the overall usability of the proposed control system.

\appendix
\section{Derivation of Eq. \ref{eq::alpha_eff}}
\label{sec:appendixA}

Given the expression \eqref{eq::F_normaliz}:
\begin{equation}
F(\mathbf{k}) = \alpha J_\mathrm{N}(\mathbf{k}) \cdot J_{\mathrm{MAX}}(\mathbf{k}) + (1-\alpha) \Gamma_\mathrm{N}(\mathbf{k}) \cdot \Gamma_{\mathrm{MAX}}(\mathbf{k}), \nonumber
\end{equation}
we want to obtain a normalized expression of the form
\begin{equation}
F(\mathbf{k}) = \alpha_{\mathrm{eq}} J_\mathrm{N}(\mathbf{k}) + (1-\alpha_{\mathrm{eq}}) \Gamma_\mathrm{N}(\mathbf{k}),
\end{equation}
in which $\alpha_{\mathrm{eq}}$ is the effective parameter of \eqref{eq::F_normaliz} when both $J(\mathbf{k})$ and $\Gamma(\mathbf{k})$ are normalized in $[0,1]$ using the respective maximum values.
First, we impose that the relative weighting provided by the user-defined $\alpha$ is preserved in the analytically calculated $\alpha_{\mathrm{eq}}$.
This is done by considering the following ratio:
\begin{equation}
\begin{aligned}
\label{eq:ratio}
&\frac{\alpha J_{\mathrm{N}}(\mathbf{k}) J_{\mathrm{MAX}}(\mathbf{k})}{(1-\alpha)\Gamma_{\mathrm{N}}(\mathbf{k})\Gamma_{\mathrm{MAX}}(\mathbf{k})} = \frac{\alpha_{\mathrm{eq}}J_{\mathrm{N}}(\mathbf{k})}{(1-\alpha_{\mathrm{eq}})\Gamma_{\mathrm{N}}(\mathbf{k})} \\
&\Rightarrow \frac{\alpha J_{\mathrm{MAX}}(\mathbf{k})}{(1-\alpha)\Gamma_{\mathrm{MAX}}(\mathbf{k}) } = \frac{\alpha_{\mathrm{eq}}}{1-\alpha_{\mathrm{eq}}}.
\end{aligned}
\end{equation}

Accordingly, we can compute $\alpha_{\mathrm{eq}}$ by the following manipulations of \eqref{eq:ratio}:
\begin{align}
\alpha_{\mathrm{eq}} &= (1-\alpha_{\mathrm{eq}}) \cdot \frac{\alpha J_{\mathrm{MAX}}(\mathbf{k})}{(1-\alpha)\Gamma_{\mathrm{MAX}}(\mathbf{k})}; \nonumber \\
\alpha_{\mathrm{eq}} &= \frac{\alpha J_{\mathrm{MAX}}(\mathbf{k})}{(1-\alpha)\Gamma_{\mathrm{MAX}}(\mathbf{k})} - \alpha_{\mathrm{eq}} \cdot \frac{\alpha J_{\mathrm{MAX}}(\mathbf{k})}{(1-\alpha)\Gamma_{\mathrm{MAX}}(\mathbf{k})}; \nonumber \\
\alpha_{\mathrm{eq}} &\cdot \frac{(1-\alpha)\Gamma_{\mathrm{MAX}}(\mathbf{k}) + \alpha \cdot J_{\mathrm{MAX}}(\mathbf{k})}{(1-\alpha)\Gamma_{\mathrm{MAX}}(\mathbf{k})} = \frac{\alpha J_{\mathrm{MAX}}(\mathbf{k})}{(1-\alpha)\Gamma_{\mathrm{MAX}}(\mathbf{k})}; \nonumber \\
\nonumber\alpha_{\mathrm{eq}} &= \frac{\alpha J_{\mathrm{MAX}}(\mathbf{k})}{(1-\alpha)\Gamma_{\mathrm{MAX}}(\mathbf{k}) + \alpha \cdot J_{\mathrm{MAX}}(\mathbf{k})}.
\end{align}

\bibliographystyle{abbrvnat}
\bibliography{Bibliography}
\end{document}